\documentstyle[aps,manuscript,epsf]{revtex} %twocolums % al reves

\begin{document}
%\twocolumn[{  %twocolums
%\draft        %twocolums
%
\title{\bf 
%%%%%%%%%%%%%%%%%%%%%%%%%%%%%%%%%%%%%%%%%%%%%%%%%%%%%%%%%%%%%%%%%%%%%%%%%%%%%%%
%%%%%%%%%%%%%%%%%%%%%%%%%%%%%%%%%%%%%%%%%%%%%%%%%%%%%%%%%%%%%%%%%%%%%%%%%%%%%%%
%                     The Truth About Earthquakes:                
%Scaling and universality in the time interval between successive earthquakes
Long-Term Clustering, Scaling, and Universality 
in the Temporal Occurrence of Earthquakes
%%%%%%%%%%%%%%%%%%%%%%%%%%%%%%%%%%%%%%%%%%%%%%%%%%%%%%%%%%%%%%%%%%%%%%%%%%%%%%%
%%%%%%%%%%%%%%%%%%%%%%%%%%%%%%%%%%%%%%%%%%%%%%%%%%%%%%%%%%%%%%%%%%%%%%%%%%%%%%%
}
\author{
%%%%%%%%%%%%%%%%%%%%%%%%%%%%%%%%%%%%%%%%%%%%%%%%%%%%%%%%%%%%%%%%%%%%%%%%%%%%%%%
%Santi Monguitoh %$^1$\cite{underg},
%and
\'Alvaro Corral%
\footnote{
%Grup de F\'\i sica Estad\'\i stica,
Departament de F\'\i sica, Facultat de Ci\`encies, 
Universitat Aut\`onoma de Barcelona,
%Edifici Cc, 
E-08193 Bellaterra, Barcelona, Spain 
}
%$^{1,2}$%$^{\dagger,}$ 
%\cite{email} 
%%%%%%%%%%%%%%%%%%%%%%%%%%%%%%%%%%%%%%%%%%%%%%%%%%%%%%%%%%%%%%%%%%%%%%%%%%%%%%%
}
%\address{
%%______________________________________________________________________________
%%%$^{\dagger}$%
%%$^1$
%%The Niels Bohr Institute, University of Copenhagen,       
%%Blegdamsvej 17,  DK-2100  Copenhagen \O,   Denmark \\
%%$^2$
%Grup de F\'\i sica Estad\'\i stica,
%Departament de F\'\i sica, %Facultat de Ci\`encies, 
%Universitat Aut\`onoma de Barcelona,
%Edifici Cc, E-08193 Bellaterra, Barcelona, Spain 
%%-------------------------------------------------------------------------------
%}
%
%_______________________________________________________________________________
%%\date{Accepted in PRE 15 July 2002; published ???}
\date{\today}
%-------------------------------------------------------------------------------

\maketitle %\parskip 2ex
%
%\widetext  %twocolums
%
\begin{abstract}
%\leftskip 54.8 pt %twocolums
%\rightskip 54.8 pt %twocolums 
%____________________________ ABSTRACT ________________________________________
%%We study one worldwide seismic catalog, as well as several regional catalogs, 
%%in order to measure the recurrence-time distribution of earthquakes. 
%%By means of scaling analysis, it is found that data 
%
%Applying scaling analysis to
Analyzing 
diverse seismic catalogs, we have determined that the
probability densities of the earthquake recurrence times %of earthquakes
for different spatial areas and magnitude ranges 
can be described by a unique universal distribution 
if the time is rescaled with the mean rate of occurrence.
%%defined as the number of events per unit time. 
%The power-law nature 
The shape of the distribution shows the existence of clustering 
beyond the short-term regime, 
and scaling reveals the self-similarity of the clustering structure. 
This holds from worldwide to local scales and for quite different
tectonic environments.
Aftershock
sequences also follow this universal recurrence-time distribution if the rescaling is
undertaken with the time-varying rate.
%%These findings are valid even when the occurrence rate is not stationary, 
%%as exemplified by aftershocks sequences.
%-------------------------------------------------------------------------------
\end{abstract}
%
%\leftskip 54.8 pt %twocolums
%
%
%}]   %twocolums

\narrowtext
%
%\setcounter{page}{1} %twocolums
%_______________________________________________________________________________
%\markright{\bf \today} %twocolums
%-------------------------------------------------------------------------------
%\thispagestyle{myheadings} %twocolums
%\pagestyle{myheadings} %twocolums
%
\newpage   %twocolums % este va al reves

%%%%%%%%%%%%%%%%%%%%%%%%%%%%%%%%%%%%%%%%%%%%%%%%%%%%%%%%%%%%%%%%%%%
%\section{ Introduction.}
%%%%%%%%%%%%%%%%%%%%%%%%%%%%%%%%%%%%%%%%%%%%%%%%%%%%%%%%%%%%%%%%%%%

%Although earthquakes are a phenomenon of great complexity, certain simple
%general laws govern the statistics of their occurrence {\it (1-5)}; 
%however, for the time
%interval between successive earthquakes, no such a general law has yet been
%established {\it (5,6)}. 
%analyzing diverse seismic catalogs, we have determined that the
%probability density of the recurrence times 
%in a given region decays for
%intermediate times as a power law, which is accelerated by an exponential factor
%for greater times. 
%The time scale is set by the mean rate of occurrence in such a
%way that distributions for different regions scale with their corresponding rates,
%collapsing onto a single universal curve under rescaling. 
%The power-law
%distribution features the existence of clustering beyond the short-term regime  {\it (7)}, and
%scaling implies the self-similarity of the clustering structure. 
%These results are valid
%from worldwide to local scales, for quite different tectonic regions, 
%for a broad
%range of magnitudes, and when using all the events in the catalogs. Aftershock
%sequences also follow this universal recurrence-time distribution if the rescaling is
%undertaken with the time-varying rate.

Although earthquakes are a phenomenon of great complexity, certain simple
general laws govern the statistics of their occurrence {\it (1-5)}; 
however, 
no such a unified description has yet been established 
for the time interval between successive events 
{\it (6-9)}. 
On the contrary, %due to
the rich variability intrinsic to earthquakes
%has hindered the search for a
%unified description, and has provoked that
has promoted that
all possibilities have been proposed,
%for the temporal behavior, 
from totally random occurrence to the periodic ticking of great quakes.
The most extended view is that of two separated processes, one for mainshocks, which
ought to follow a Poisson distribution {\it (10)} (or not {\it (6-9)}), 
and an independent process to generate aftershocks. 
% (and foreshocks). 
Consequently, the ``standard practice'' for this approach
consists first of delimiting the spatial area to be studied, on the basis of its
tectonic characteristics, and then of a careful (and not so standard) identification of
aftershocks, in order to separate them from the main sequence.

Here we take an alternative perspective, complementary to the previous reductionist
view. 
%Inspired by complex-system research, 
We try to look at the system as a whole,
irrespective of tectonic features and placing all the events 
on the same footing, whether
these would be classified as mainshocks or aftershocks {\it (11,12)}. 
This follows one of the key
guidelines of complexity philosophy, 
which is to find descriptions on a general level;
the existence of general laws fulfilled by all the earthquakes 
will unveil a degree of
unity in an extremely complex phenomenon {\it (13)}.

We analyze a global catalog, the PDE from the NEIC {\it (14)}, as well as several local
catalogs: that of the SCSN (Southern California) {\it (15)}, 
the JUNEC (Japan) {\it (16)}, 
the Bulletins of the IGN (the Iberian Peninsula and the North of Africa) {\it (17)},
and the BGS catalog (the British Islands and the North Sea) {\it (18)}. 
Catalogs generally characterize
each earthquake by three main quantities: time of occurrence, magnitude, and a vector of
spatial coordinates for the hypocenter. These are also the variables that we focus on.

Without concerning ourselves with the tectonic properties, following Bak {\it et al.}
{\it (11,12)}, 
we consider spatial areas delimited by a window of $L$ degrees in longitude and $L$
degrees in latitude
(this corresponds to a square region if these angles are translated onto
a rectangular coordinate system {\it (19)}). 
%nevertheless, the shape of the region is totally
%irrelevant in our study). 
For each one of these regions, only events with magnitude $M$
above a certain threshold $M_c$ are considered 
(the threshold should be larger than the
minimum magnitude for which the catalog is considered complete).
%, this value can be
%tested by the fulfillment of the Gutenberg-Richter relation {\it (1,2)}). 
In this way, we transform
a time process in four dimensions (spatial coordinates and magnitude) into a simple
process on a line for which events occur at times 
$t_i$, with $i=1,2 \dots N$, 
and therefore, the
time between successive events can be obtained as 
$
\tau_i = t _{i+1}  - t_i,  i=1,2 \dots N-1.
$
These are the recurrence times in a given $L^2-$region for events above $M_c$, which can
also be referred to as inter-occurrence or waiting times. Note that, with this
transformation, we have lost the structure in space and in the magnitude scale;
nonetheless, the change in the process properties with the variation of $L$ and $M_c$  
will allow us to recover some of this information.

Due to the multiple time scales involved (from seconds or minutes to many years),
the probability density of the recurrence time must be calculated with care. We could
work with the logarithm of $\tau$, but an equivalent and more direct possibility is to define
the bins over which the probability density is calculated exponentially growing as $c^n$, 
with $c > 1$ and $n$ labeling consecutive bins (we usually take $c = 2.5$). 
This ensures that
we have the appropriate bin size for each time scale. We then count the number of pairs
of consecutive events separated by a time whose value lies into a given bin, and divide
by the total number of pair of events (number of events minus one) and by the size of
the bin, to attain the estimation of the probability density $D_{xy}(\tau)$
over that bin, where $xy$ denotes the spatial coordinates of the region. 
($D_{xy}$ also depends on $L$ and $M_c$, but for the
sake of simplicity in the notation, we obviate this dependence.) 
Moreover, due to the incompleteness of the catalogs
in the short-time scale, we will not display in the plots
time intervals that are smaller than 2 minutes.
%Moreover, in order to
%avoid errors in time location and the ``shadow effect'' 
%that earthquakes produce in the
%recording instruments when they overlap, 
%we disregard time intervals that are smaller
%than 2 minutes.

The entire Earth has been analyzed by this method. Figure 1A shows the results for
$D_{xy}(\tau)$  for worldwide earthquakes in the NEIC-PDE catalog for the 1973-2002
period, using $L$ from $180^\circ$ to $2.8^\circ$ (about 300 km) and $M_c$  from 5 to 6.5. 
Note the
variation of the recurrence time across several orders of magnitude. Figure 1B shows the
rescaling of all the distributions with the mean rate $R$ in the region, defined as the total
number of events divided by the total time interval over which these events span. The
perfect data collapse implies that we can write
$$
D_{xy}(\tau)=R_{xy} f(R_{xy} \tau),
$$
where $R_{xy}$ stresses that the rate refers to the $(x,y)$-region 
(of size $L^2$ and with $M \ge M_c$).
The scaling function $f$ can be well fit by a generalized gamma distribution,
$$
    f(\theta) = 
           C \frac{1}{\theta^{1-\gamma}}
           \exp(-\theta^\delta/B),
$$
with parameters $\gamma= 0.67 \pm 0.05$, $\delta  = 0.98 \pm 0.05$, 
$B = 1.58 \pm 0.15$, and $C = 0.50 \pm 0.10$,
which has a coefficient of variation $CV \simeq 1.2$. 
In fact, the value of delta can be
approximated to one, which corresponds to the standard gamma distribution; we
therefore essentially have a power law with exponent about $-0.33$, up to the largest
values, where the exponential factor comes into play.

The scaling function fits the rescaled distributions surprisingly well for
intermediate and large values of the recurrence time, about 
$\tau > 0.01/R_{xy}$,
(this usually
contains from $90 \%$ to $95 \%$  of probability). The deviations are considerable for small
values of $\tau$; although the statistic is low in this case (few events in the small bins being
considered), for certain regions there is a clear tendency for the distribution to exceed
the value given by the scaling function, that is, there is an excess of very short
recurrence times, in the form of another power law but with the exponent close to
(minus) one. This occurs when the rate in the region is not stationary, due to the sudden
increase and slow decay of the activity provoked by aftershock sequences. Furthermore,
these increments become more apparent when the size of the region decreases, in such a
way that, for $L \le 11^\circ$, not all the regions in the world verify the scaling law; this can be
solved in some cases by rescaling with the mean rate in the region calculated, not over
the whole time-span of the catalog, but only over the period for which the rate is
stationary (no activity peaks). 
Nevertheless, the aftershocks can be so important for
certain particular regions that a stationary period may not exist.

The same analysis is performed on local catalogs and identical results hold, as
Fig. 1C illustrates. For Southern California for the $1984-2001$ period, a number of small
regions with stationary activity are shown; for larger regions, the time window must be
reduced to $1988-1991$ or to $1995-1998$, for instance, in order to find stationariness. For
Japan, we also analyze large regions for the $1995-1998$ period, for the Iberian Peninsula
the period is $1993-1997$, and for the British Islands, $1991-2001$. The magnitude
thresholds range from $2$ to $4$, and $L$ from $30^\circ$ to $0.16^\circ$ 
(approximately $3300$ to $17$ km).

The shape of the distribution $D_{xy}(\tau)$
indicates that the memory of the last
earthquake is conserved up to the largest times, with the probability of a subsequent
event being maximum immediately after the last shock, and slowly decreasing with
time. This constitutes a clustering effect {\it (20)}, 
in which earthquakes attract each other, and
has as a counter-intuitive consequence the fact that the longer it has been since the last
earthquake, the longer the expected time will be till the next {\it (21)}. 
On the other hand, the scaling of $D_{xy}(\tau)$
under changes in $M_c$, $L$, and the region coordinates implies that the
clustering structure is self-similar over different regions and magnitude ranges. The
robustness of the distribution under such changes is therefore noteworthy. 
It is also
remarkable that if the region is kept fixed and only $M_c$  varies, the scaling with the rate
$R_{xy}$ can be substituted by the factor $10^{-b M_c}$, where $b$ refers to the $b-$value of the
frequency-magnitude relation {\it (1,2)} in that particular region {\it (11,12)}; 
despite the regional variability of $b$, 
the universality of the scaling function $f$ remains valid.

At the outset of this exposition, we suggested that our results were valid for all
events, including aftershocks; however, the aftershocks can break the scaling of the
distribution up to very large time values. We can then say that the universal distribution
does not describe the short-time intervals over which aftershocks replicate. This may
appear to be correct, but can be turn out to be mistaken in the following way: for
aftershock sequences, the mean rate is not stationary, but changes with time; therefore,
we should rescale the recurrence times using the ``instantaneous'' rate $r_{xy}(t)$. 
For many sequences, $r_{xy}(t)$ is found to decay following the Omori law (3),
$$
%r_{xy}(t)=\frac{A}{(c+t)^p}
r_{xy}(t)=\frac{A}{t^p}
$$
where $t$ is the time elapsed since the mainshock (and the parameters depend on $x, y, L,$
and $M_c$). Figure 2A shows precisely this for several important earthquakes in Southern
California {\it (15)}. The corresponding recurrence-time distributions, calculated over the period
for which the Omori law is fulfilled, are displayed in Fig. 2B; the results after rescaling
with $r_{xy}(t)$ appear in Fig. 2C, again in agreement with the universal distribution. 
The scaling is outstanding, taking into account that the $p-$values
spread from 0.9 to 1.35.
Note also that this scaling 
implies the existence of a secondary clustering inside the primary clustering of
the aftershock sequence,
and therefore the process is not a nonhomogeneous Poisson process.

The present characterization of the stochastic spatio-temporal
occurrence of earthquakes by means of a unique law 
would indicate the existence of universal mechanisms
in the earthquake-generation process {\it (22)}, 
the understanding of which, however, is still far beyond us.
Nevertheless, the context of self-organized critical phenomena 
{\it (11-13)} provides a coherent framework at this stage.
These findings can also be relevant to
continuous-time random-walk models of seismicity {\it (23,24)}, 
time-dependent hazard, and forecasting in general {\it (25-28)}.

%\begin{references}
%_________________________________________________________________________
%                                 REFERENCES
%-------------------------------------------------------------------------
%

%\bibitem[*]{email}
%E-mail address:
%{\tt Alvaro.Corral@uab.es}

%\begin{center}
\vspace{1cm}
\phantom{1. 1. }{\bf References and Notes}
%\end{center}

\begin{enumerate}

\item
%1. 
B. Gutenberg, C. F. Richter, 
{\it Seismicity of the Earth}
(Hafner Pub. Co., New York, 1965).

\item %2. 
D. L. Turcotte, 
{\it Fractals and Chaos in Geology and Geophysics }
(Cambridge University Press, Cambridge, ed. 2, 1997).

\item %3. 
T. Utsu, Y. Ogata, R. S. Matsu'ura, 
%The centenary of the Omori formula for a decay law of aftershock activity. 
{\it J. Phys. Earth }{\bf 43}, 1 %-33 
(1995).

\item 
P. A. Reasenberg, L. M. Jones,
%Earthquake hazard after a mainshock in California.
{\it Science} {\bf 243}, 1173 %-1176
(1989).

\item %4. 
Y. Y. Kagan, L. Knopoff,
%Spatial distribution of earthquakes: the two-point correlation function. 
{\it Geophys. J. R. astr. Soc.} {\bf 62}, 303 %-320 
(1980).

\item %5. 
D. Sornette, L. Knopoff,
%The paradox of the expected time until the next earthquake. 
{\it Bull. Seism. Soc. Am.} {\bf 87}, 789 %-798 
(1997).

\item %6. 
J.-H. Wang, C.-H. Kuo,
%On the frequency distribution of interoccurrence times of earthquakes. 
{\it J. Seism.} {\bf 2}, 351 %-358 
(1998).

\item %7.
W. L. Ellsworth {\it et al.},
``A physically-based earthquake recurrence model 
for estimation of long-term earthquake probabilities''
(U.S. Geological Survey Open-File Report 99-522, 1999). 

\item
R. F. Smalley, J.-L. Chatelain, D. L. Turcotte, R. Pr\'{e}vot,
%A fractal approach to the clustering of earthquakes:
%applications to the seismicity of the New Hebrides.
{\it Bull. Seism. Soc. Am.} {\bf 77}, 1368 %-1380?
(1987).

\item %8. 
J. K. Gardner, L. Knopoff,
%Is the sequence of earthquakes in Southern California,
%with aftershocks removed, Poissonian? 
{\it Bull. Seism. Soc. Am.} {\bf 64}, 1363 %-1367 
(1974).

\item %9. 
P. Bak, K. Christensen, L. Danon, T. Scanlon,
%Unified scaling law for earthquakes. 
{\it Phys. Rev. Lett.} {\bf 88}, 178501 (2002).

\item %10. 
K. Christensen, L. Danon, T. Scanlon, P. Bak,
%Unified scaling law for earthquakes. 
{\it Proc. Natl. Acad. Sci. USA} {\bf 99}, 2509 %-2513 
(2002).

\item %11. 
P. Bak, 
{\it How Nature Works: The Science of Self-Organized Criticality }
(Copernicus, New York, 1996).

\item %12. 
National Earthquake Information Center, Preliminary Determination of Epicenters
catalog, 
http://wwwneic.cr.usgs.gov/neis/epic/epic$\_$global.html.

\item %13. 
Southern California Seismographic Network,
http://www.scecdc.scec.org/ftp/catalogs/SCSN.

\item %14. 
Japan University Network Earthquake Catalog, 
http://wwweic.eri.u-tokyo.ac.jp/CATALOG/junec/monthly.html.

\item %15. 
Instituto Geogr\'{a}fico Nacional, Bolet\'{\i}n de Sismos Pr\'{o}ximos,
http://www.geo.ign.es/servidor/sismo/cnis/terremotos.html.

\item %16. 
British Geological Survey, catalog available upon request.

\item %17. 
Nevertheless, the shape of the region is totally
irrelevant in our study.

\item %18 
Y. Y. Kagan, D. D. Jackson,
%Long-term earthquake clustering. 
{\it Geophys. J. Int.} {\bf 104}, 117 %-133 
(1991).

\item 
P. M. Davis, D. D. Jackson, Y. Y. Kagan,
%The longer it has been since the last earthquake, 
%the longer the expected time till the next? 
{\it Bull. Seism. Soc. Am.} {\bf 79}, 1439 %-1456 
(1989).

\item % 19
S. Toda, R. S. Stein, T. Sagiya,
%Evidence from the AD 2000 Izu islands earthquake swarm
%that stressing rate governs seismicity.
{\it Nature} {\bf 419}, 58 %-61
(2002).

\item %
Y. Ogata,
%Seismicity analysis through point-process modeling: a review. 
{\it Pure appl. geophys.} {\bf 155}, 471 %-507 
(1999).

\item %
A. Helmstetter, D. Sornette,
%Diffusion of epicenters of earthquake aftershocks, Omori's law, 
%and generalized continuous-time random walk models. 
{\it Phys. Rev. E} {\bf 66}, 061104 (2002).

\item % 
I. Main,
%Long odds on prediction. 
{\it Nature} {\bf 385}, 19 %-20 
(1997).

\item %
R. J. Geller, D. D. Jackson, Y. Y. Kagan, F. Mulargia,
%Earthquakes cannot be predicted. 
{\it Science} {\bf 275}, 1616 %-1617 
(1997).

\item %
Nature Debates, 
``Is the reliable prediction of individual earthquakes a realistic
scientific goal?''
http://www.nature.com/nature/debates/index.html.

\item % 
T. Parsons, S. Toda, R. S. Stein, A. Barka, J. H. Dieterich,
%Heightened odds of large earthquakes near Istanbul: 
%an interaction-based probability calculation. 
{\it Science} {\bf 288}, 661 %-665 
(2000).

\item %
This work would have been impossible without the fundamental 
ideas put forward by the late Per Bak.
The author is also grateful to M. Bogu\~n\'a, K. Christensen, 
D. Pav\'on, the Ram\'on y Cajal program, and all the people
at the Statistical Physics Group of the Universitat Aut\`onoma de Barcelona, 
as well as to those institutions that have
made their catalogs available on the Internet.

\end{enumerate}

%\end{references}

%%}] %twocolums

\newpage

\vspace{1cm}
{\bf Figure Captions}

{\bf Fig. 1.} Recurrence-time distributions without and with rescaling. 
{\bf (A)} Probability
densities from the NEIC-PDE worldwide catalog for several regions, $L$, and
$M_c$. For $L \ge 45^\circ$, all the regions with more than 500 events are shown, whereas
for $L \le 22.5^\circ$, only representative regions with moderate aftershock activity are
displayed. The vector $(k_x, k_y)$ labels the different regions, for which the
coordinates of the center can be obtained as 
$x = x_{min} + (k_x + 0.5) L$, 
$y = y_{min} + (k_y + 0.5) L$, with 
$x_{min} = -180^\circ$, 
$y_{min} =  -90^\circ$. 
(The $360^\circ \times 180^\circ$ region, which covers
the whole planet, has been included for completeness.) 
{\bf (B)} Previous data, after
rescaling, with a fit of the scaling function $f$. 
{\bf (C)} Rescaled distributions from local
catalogs. SC88, SC95, SC84 refer to Southern California for the years $1988-
1991$, $1995-1998$, and $1984-2001$. To obtain region coordinates, use the
previous formula with 
$(x_{min}, y_{min}) = (-124^\circ, 29^\circ), (-123^\circ, 30^\circ), 
(125^\circ, 25^\circ), (-20^\circ, 30^\circ)$, and
$(-10^\circ, 45^\circ)$ for SC88-95, SC84, Japan, the Iberian Peninsula, and the British Islands
respectively. 
The function displayed is the fit obtained from the NEIC-PDE catalog.

{\bf Fig. 2.} Analysis of aftershock sequences. 
{\bf (A)} Decay of the rate after a
mainshock and illustration of the Omori law for the following earthquakes in
Southern California: Chalfant Valley (July 21, 1986, $M = 5.9$), Landers (June 28,
1992, $M = 7.3$), Northridge (Jan. 17, 1994, $M = 6.7$), and Hector Mine (Oct. 16,
1999, $M = 7.1$). Regions of diverse size $L$ are considered, all of these including
the mainshock. Some curves are shifted for the sake of clarity. 
{\bf (B)} Distributions
of recurrence times for the previous sequences. 
{\bf (C)} Distributions $\psi$ of the
dimensionless time $r_{xy}(t) \tau$, 
in total agreement with the universal scaling function $f$.

%
%########################################################################
\begin{figure}
\epsfxsize=4.5truein \hskip .5truein
\epsffile{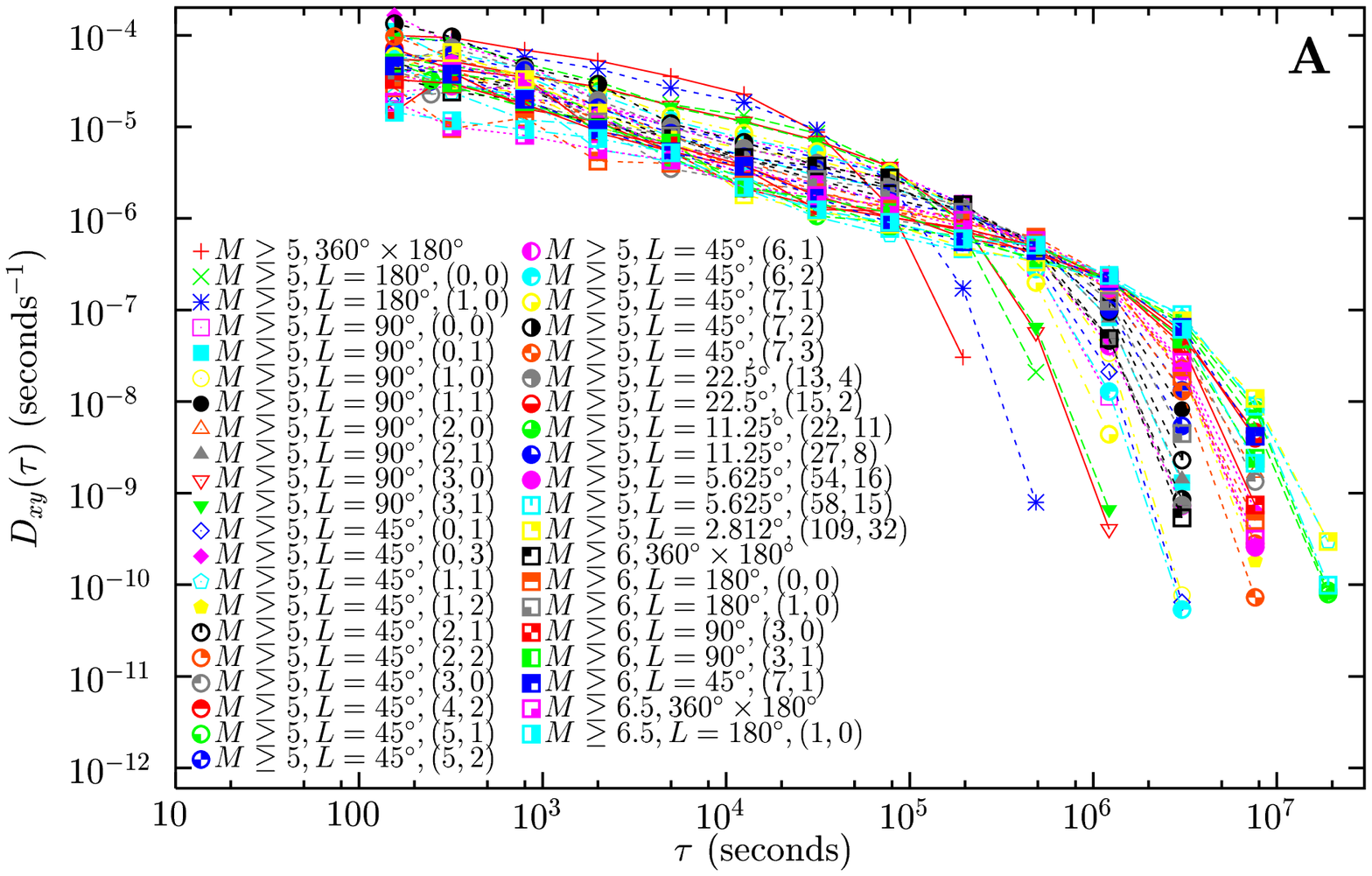} 
%\caption{Figure 1a}
\end{figure}
\begin{figure}
\vskip -1cm
\epsfxsize=4.5truein \hskip .5truein
\epsffile{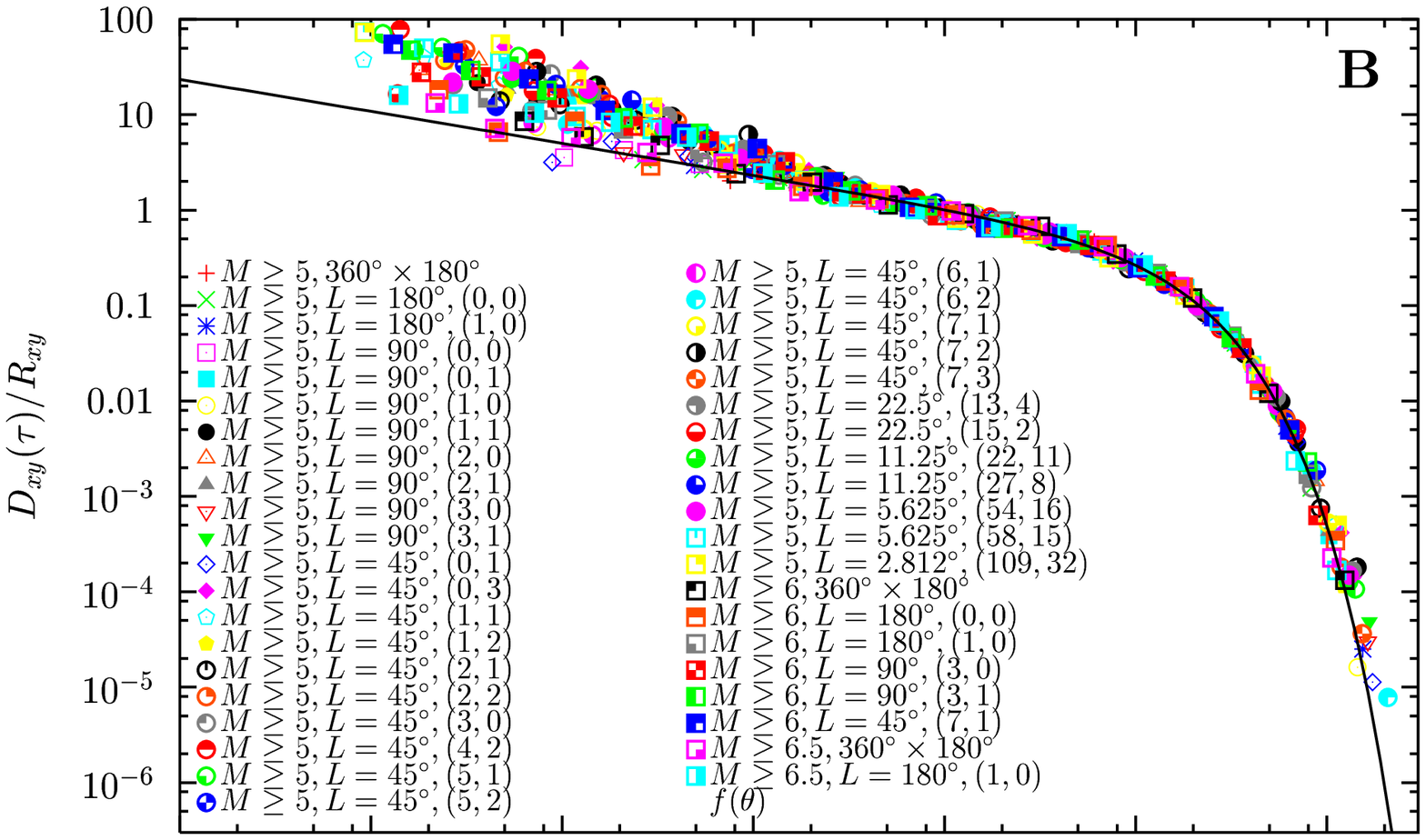} 
\end{figure}
\begin{figure}
\vskip -1.5cm
\epsfxsize=4.5truein \hskip .5truein
\epsffile{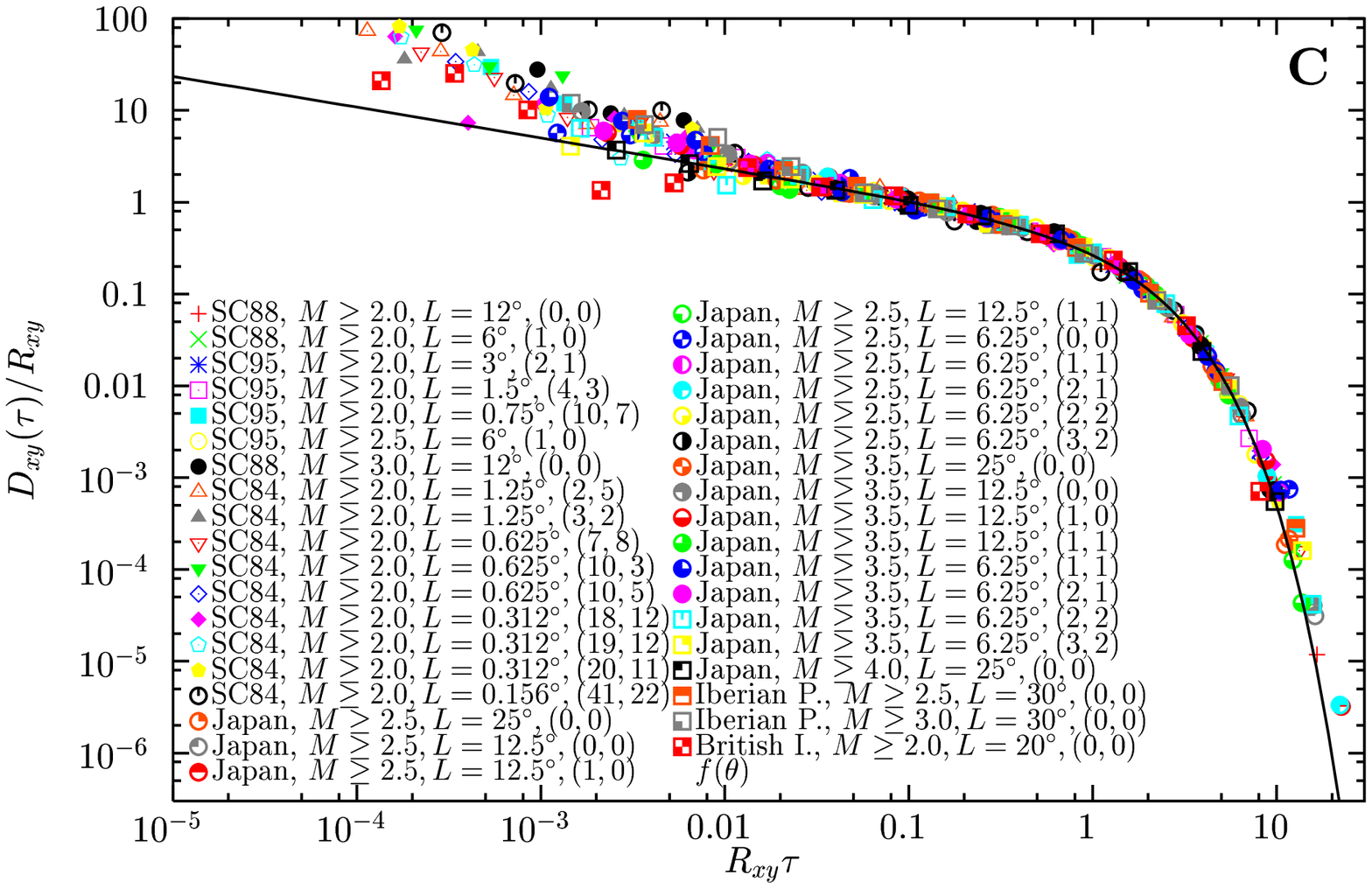} 
\caption{}
\end{figure}
%########################################################################
%
%\nopagenumbers
%
%########################################################################
\begin{figure}
\epsfxsize=4.5truein \hskip .5truein
\epsffile{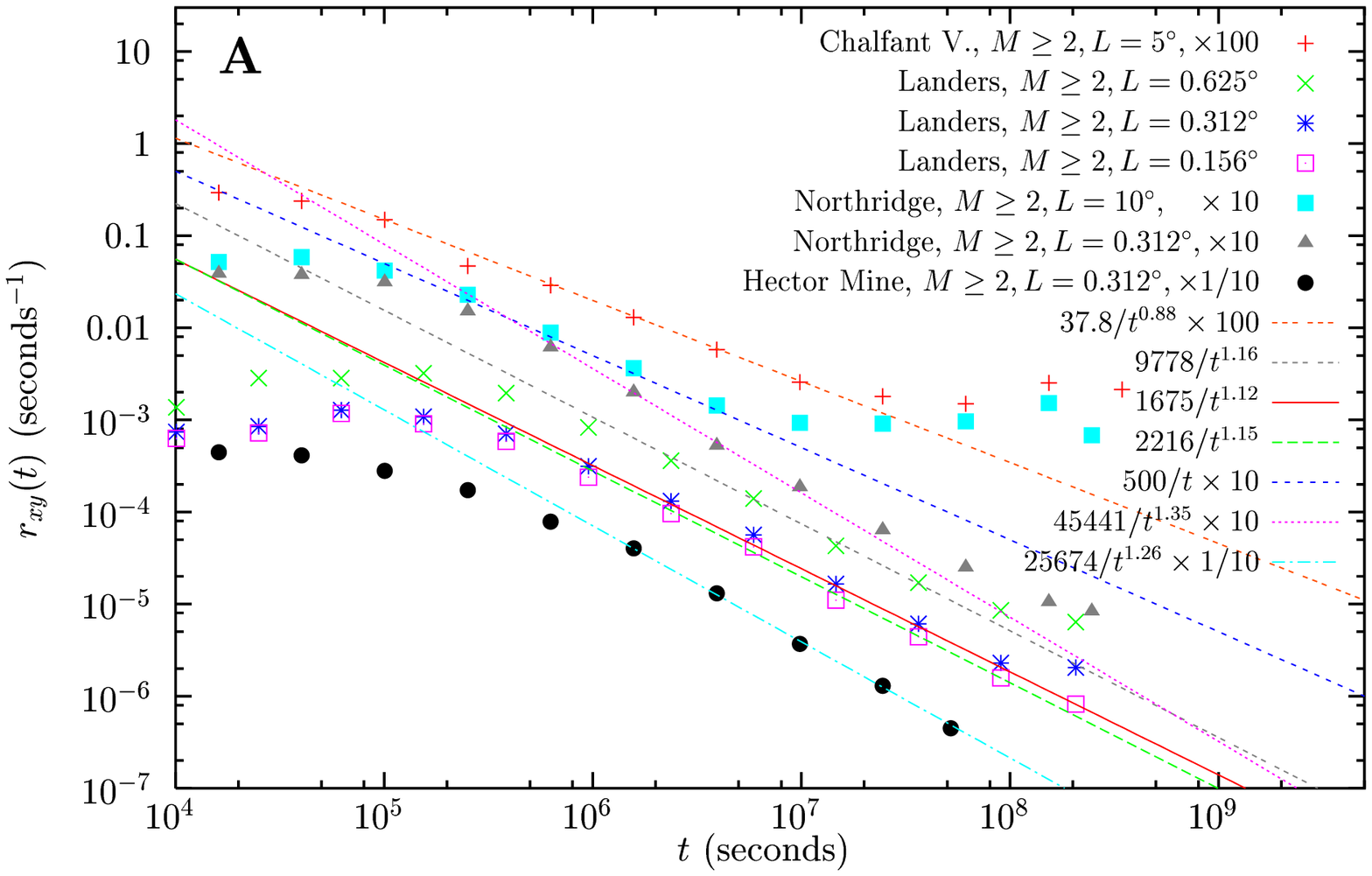} 
%\caption{Figure 1a}
\end{figure}
\begin{figure}
\vskip -1cm
\epsfxsize=4.5truein \hskip .5truein
\epsffile{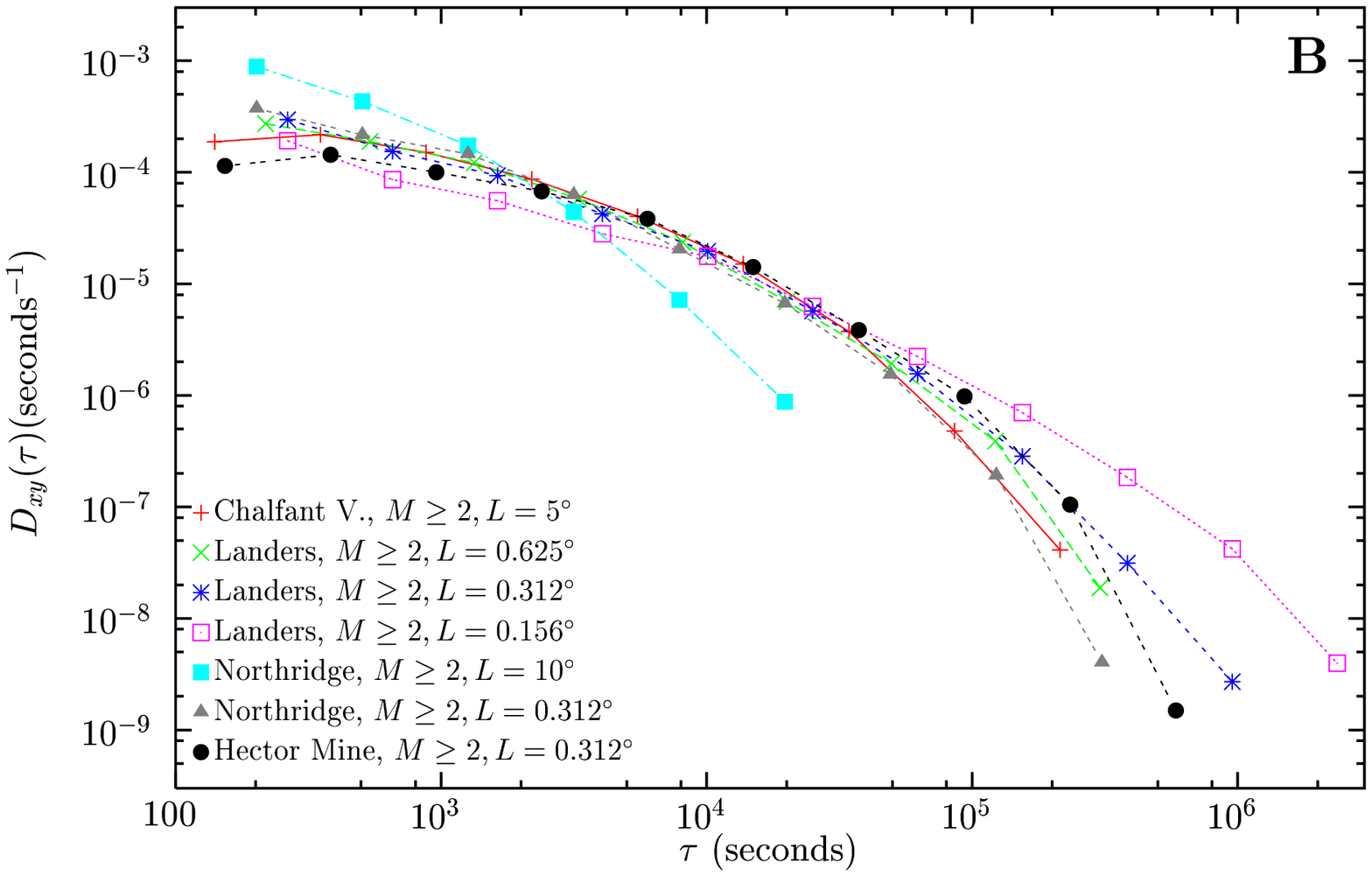} 
\end{figure}
\begin{figure}
\vskip -1cm
\epsfxsize=4.5truein \hskip .5truein
\epsffile{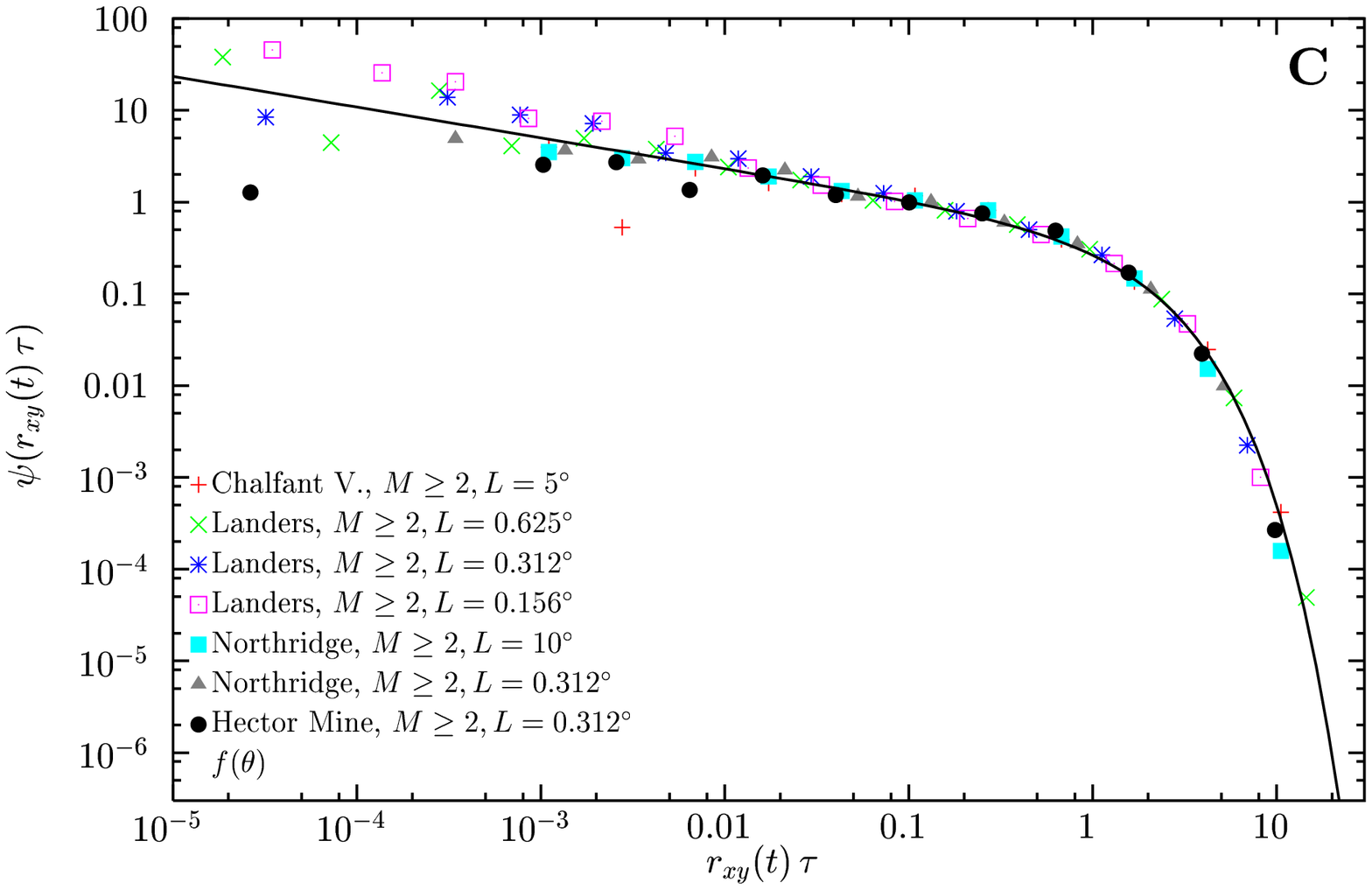} 
\caption{}
\end{figure}
%########################################################################
%

\end{document}